\documentclass[10pt]{article}
\usepackage{epsfig}
\usepackage{graphics}
\usepackage{subfigure}

\begin{document}

\begin{center}

{\large
{\bf Optimal shapes of compact strings} } \\

\vskip 2.0cm {\sc Amos Maritan$^{\ddag}$, Cristian
Micheletti$^{\ddag}$, Antonio Trovato$^{\ddag}$ \& Jayanth
R. Banavar$^{\dag}$ }

\vskip 2.0cm

\end{center}

$^{\ddag}$ International School for Advanced Studies (S.I.S.S.A.), 
Via Beirut 2-4, 34014 Trieste, Istituto Nazionale di Fisica della
Materia and the Abdus Salam International Center for Theoretical
Physics, Trieste, Italy \\

$^{\dag}$ Department of Physics and Center for Materials Physics,
104 Davey Laboratory, The Pennsylvania State University, University
Park, Pennsylvania 16802, USA \\


\vspace{1.5truecm}

{\bf Optimal geometrical arrangements, such as the stacking of atoms,
are of relevance in diverse disciplines
\cite{Sloane,Mackenzie,Wood,Car,Cipra}. A classic problem is the
determination of the optimal arrangement of spheres in three
dimensions in order to achieve the highest packing fraction; only
recently has it been proved \cite{Sloane,Mackenzie} that the answer
for infinite systems is a face-centred-cubic lattice.  This simply
stated problem has had a profound impact in many areas
\cite{Wood,Car,Cipra}, ranging from the crystallization and melting of
atomic systems, to optimal packing of objects and subdivision of
space.  Here we study an analogous problem--that of determining the
optimal shapes of closely packed compact strings. This problem is a
mathematical idealization of situations commonly encountered in
biology, chemistry and physics, involving the optimal structure of
folded polymeric chains. We find that, in cases where boundary effects
\cite{Stewart} are not dominant, helices with a particular
pitch-radius ratio are selected. Interestingly, the same geometry is
observed in helices in naturally-occurring proteins.}  \\

\newpage

The problem of placing spheres in three dimensions in order to attain
the highest density was first posed by Kepler and has attracted much
interest culminating in its recent rigorous mathematical solution
\cite{Sloane}. The close-packed hard sphere problem may be restated in
an alternative manner, more convenient for numerical implementation,
as the determination of the arrangement of a set of points in a given
volume that results in the minimum distance between any pair of points
being as large as possible \cite{Stewart}.  It is notable that the
resulting `bulk' optimal arrangement exhibits translational invariance
in that, far from the boundaries, the local environment is the same
for all points. \\

In this letter, we introduce a new problem pertaining to the optimal
shapes of compact strings. Consider a string (an open curve) in three
dimensions. We will utilize a geometric measure \cite{Buck1} of the
curve, the `rope-length', defined as the arc length measured in units
of the thickness, which has proved to be valuable in applications of
knot theory \cite{Buck1,Knots1,Knots2,Thick3,Buck2,Cantarella}.  The
thickness $\Delta$ denotes the maximum radius of a uniform tube with
the string passing through its axis, beyond which the tube either
ceases to be smooth, owing to tight local bends, or it
self-intersects. Our focus is on finding the optimal shape of a string
of fixed arc length, subject to constraints of compactness, which
would maximize its thickness, or equivalently minimize its rope
length. \\

Following the approach of Gonzalez and Maddocks \cite{Thick3}, who
studied knotted strings, we define a global radius of curvature as
follows. The global radius of curvature of the string at a given point
is computed as the minimum radius of the circles going through that
point and all other pairs of points of the string. It generalizes the
concept of the local radius of curvature (the radius of the circle
which locally best approximates the string) by taking into account
both local (bending of the string) and non-local (proximity of another
part of the string) effects. For discretized strings the local radius
of curvature at a point is simply the radius of the circle going
through the point and its two adjoining points. The minimum of all the
global radii then defines the thickness, i.e. the minimum radius of
the circles going through any triplet of discrete points. This
coincides with the previous definition in the continuum limit,
obtained on increasing the number of discretized points (assumed to be
equally spaced) on the string keeping the string length fixed
\cite{Thick3}. \\

We used several different boundary conditions to enforce the
confinement of the string.  The simplest ones discussed here are the
confinement of a string of length $l$ within a cube of side $L$ or
constraining it to have a radius of gyration (which is the
root-mean-square distance of the discretized points from their centre
of mass) that is less than a pre-assigned value $R$. Even though
different boundary conditions influence the optimal string shape, the
overall features are found to be robust. Examples of optimal shapes,
obtained from numerical simulations, for different ratios of $l/L$ and
$l/R$ are shown in Fig. 1. In both cases, two distinct families of
strings, helices and saddles, appear.  The two families are close
competitors for optimality and different boundary conditions may
stabilize one over the other.  For example, if optimal strings of
fixed length are constrained to have a radius of gyration less than
$R$, then upon decreasing $R$, the string goes from a regime where the
trivial linear string is curled into an arc, then into a portion of
helix and finally into a saddle. When the string is constrained to lie
within a cube of size $L$, as $L$ decreases first saddles are observed
and then helices.  \\

We have also been able to find bulk-like solutions which are not
influenced by boundary effects.  Such solutions can be obtained by
imposing uniform local constraints along the string. On imposing a
minimum local density on successive segments of the string (for
example, constraining each set of six consecutive beads to have a
radius of gyration that is less than a preassigned value $R$), we
obtained perfectly helical strings, as in Fig. 2, confirming that this
is the optimal arrangement.  Note that, in close analogy with the
sphere-packing problem, the optimal shape displays translational
invariance along the chain. In all cases, the geometry of the chosen
helix is such that there is an equality of the local radius of
curvature (determined by the local bending of the string) and the
radius associated with a suitable triplet of non-consecutive points
lying in two successive turns of the helix. This is a feature that is
observed only for a special ratio $c^{*}$ of the pitch, $p$, and the
radius, $r$, of the circle projected by the helix on a plane
perpendicular to its axis.  When $p/r>c^{*}=2.512$ the global radius
of curvature is equal to the local radius with the helix thickness
given by $\Delta= r (1+p^2/(2\pi r)^2)$.  If $p/r<c^{*}$, the global
radius of curvature is strictly lower than the local radius, and the
helix thickness is determined basically by the distance between two
consecutive helix turns: $\Delta\simeq p/2$ if $p/r\ll1$.  Optimal
packing selects the very special helices corresponding to the
transition between the two regimes described above. A visual example
is provided by the optimal helix of Fig. 2.\\

For discrete strings, the critical ratio $p/r$ depends on the
discretization level.  A more robust quantity is the ratio $f$,
averaged over all the points of the string, of the minimum radius of
the circles going through each point and any two non-adjacent points
and the local radius.  For discretized strings, $f=1$ just at the
transition described above, whereas $f>1$ in the `local' regime and
$f<1$ in the `non-local' regime. In our computer-generated optimal
strings, $f$ differed from unity by less than a part in a thousand.
\\

It is interesting to note that, in nature, there are many instances of
the appearance of helices. For example, many biopolymers such, as
proteins and enzymes, have backbones which frequently form helical
motifs.  (Rose and Seltzer \cite{Rose} have used the local radii of
curvature of the backbone as input in an algorithm for finding the
peptide chain turns in a globular protein.)  It has been shown
\cite{M0} that the emergence of such motifs in proteins (unlike in
random heteropolymers which, in the melt, have structures conforming
to gaussian statistics) is the result of the evolutionary pressure
exerted by nature in the selection of native state structures that are
able to house sequences of amino acids which fold reproducibly and
rapidly \cite{M} and are characterized by a high degree of
thermodynamic stability \cite{SSK}.  Furthermore, because of the
interaction of the amino acids with the solvent, globular proteins
attain compact shapes in their folded states. \\

It is then natural to measure the shape of these helices and assess if
they are optimal in the sense described here. The measure of $f$ in
$\alpha$-helices found in naturally-occurring proteins yields an
average value for $f$ of $1.03 \pm 0.01$, hinting that, despite the
complex atomic chemistry associated with the hydrogen bond and the
covalent bonds along the backbone, helices in proteins satisfy optimal
packing constraints (for the measure of $f$ we considered
$\alpha$-helices extracted from the unrelated proteins 1erv, 1beo and
2end in the Protein Data Bank). This result implies that the backbone
sites in protein helices have an associated free volume distributed
more uniformly than in any other conformation with the same
density. This is consistent with the observation \cite{M0} that
secondary structures in natural proteins have a much larger
configurational entropy than other compact conformations. This
uniformity in the free volume distribution seems to be an essential
feature because the requirement of a maximum packing of backbone sites
by itself does not lead to secondary structure formation
\cite{ChDi,On}. Furthermore, the same result also holds for the
helices appearing in the collagen native state structure, which have a
rather different geometry (in terms of local turn angles, residues per
turn and pitch \cite{Creighton}) from average $\alpha$-helices. In
spite of these differences, we again obtained an average $f = 1.01 \pm
0.03$ (Fig. 3), very close to the optimal situation.


{\bf Acknowledgements} This work was supported by INFN, NASA and The
Donors of the Petroleum Research Fund administered by the American
Chemical Society.  A.T. thanks the Physics Department of Universit\`a
degli Studi di Padova, Padova, Italy, for its hospitality.\\



\begin{figure}
\centering
\subfigure[]{\includegraphics[width=1.2in]{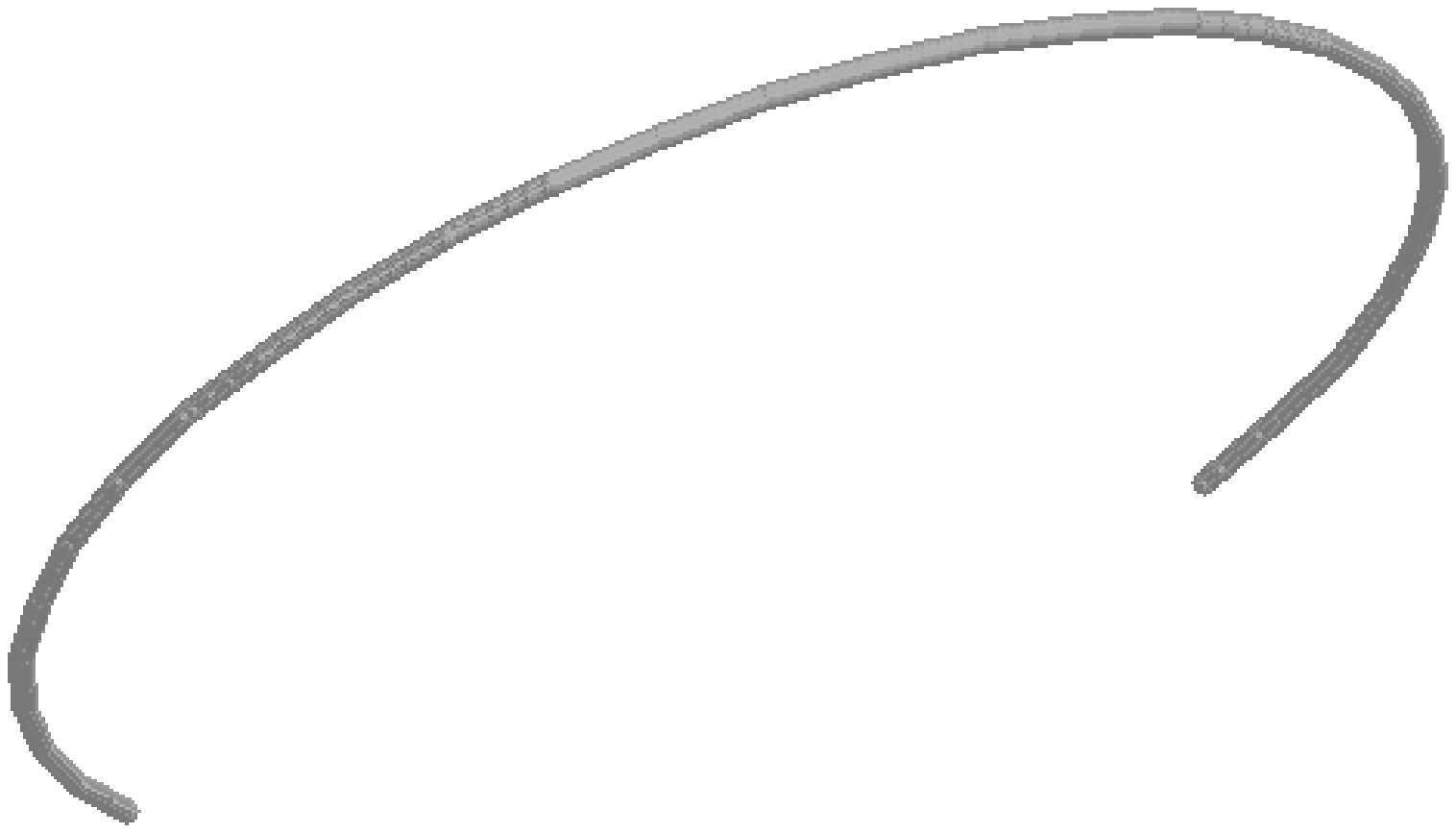}}
\hfill
\subfigure[]{\includegraphics[width=1.2in]{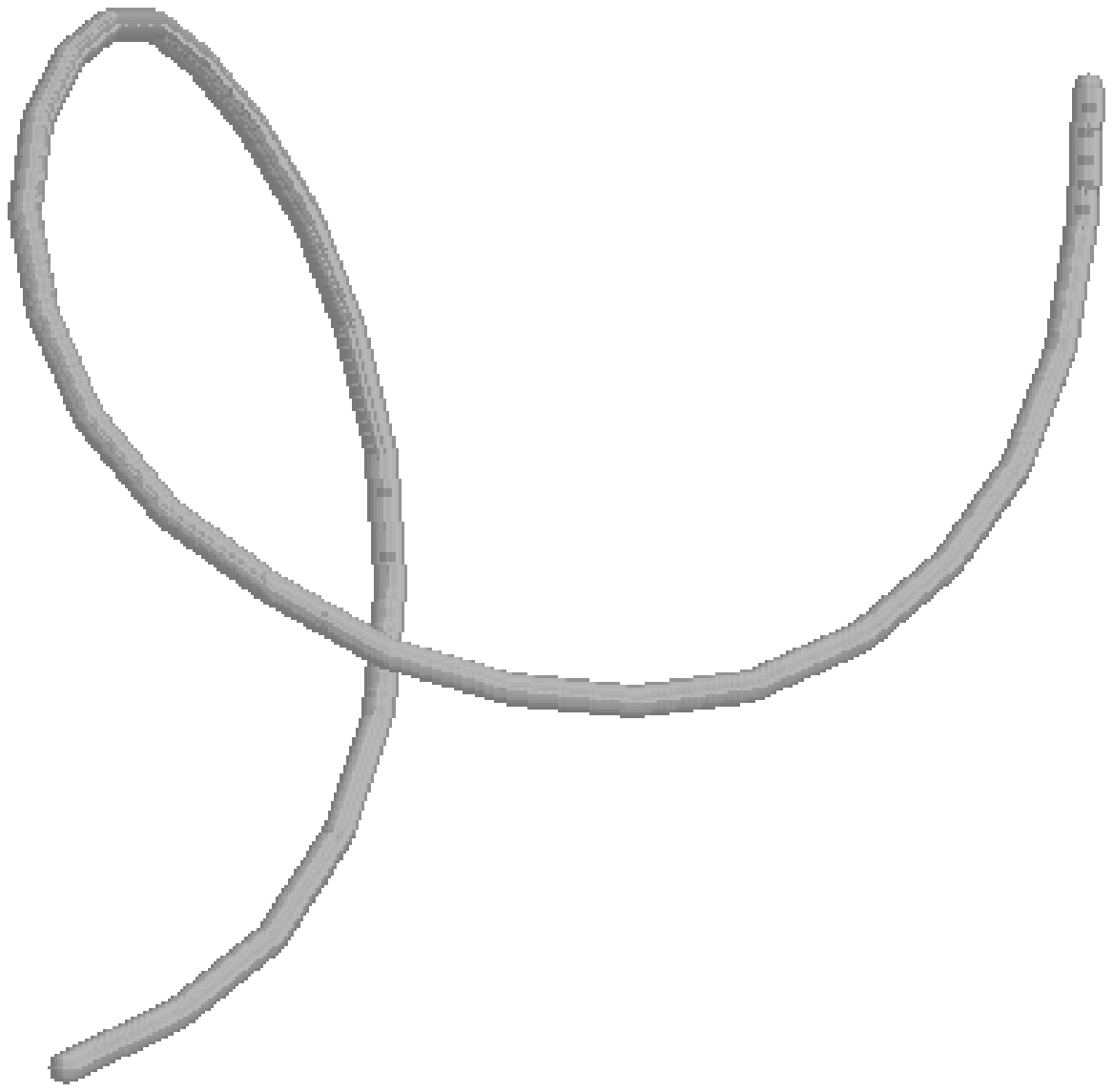}}
\hfill
\subfigure[]{\includegraphics[width=1.2in]{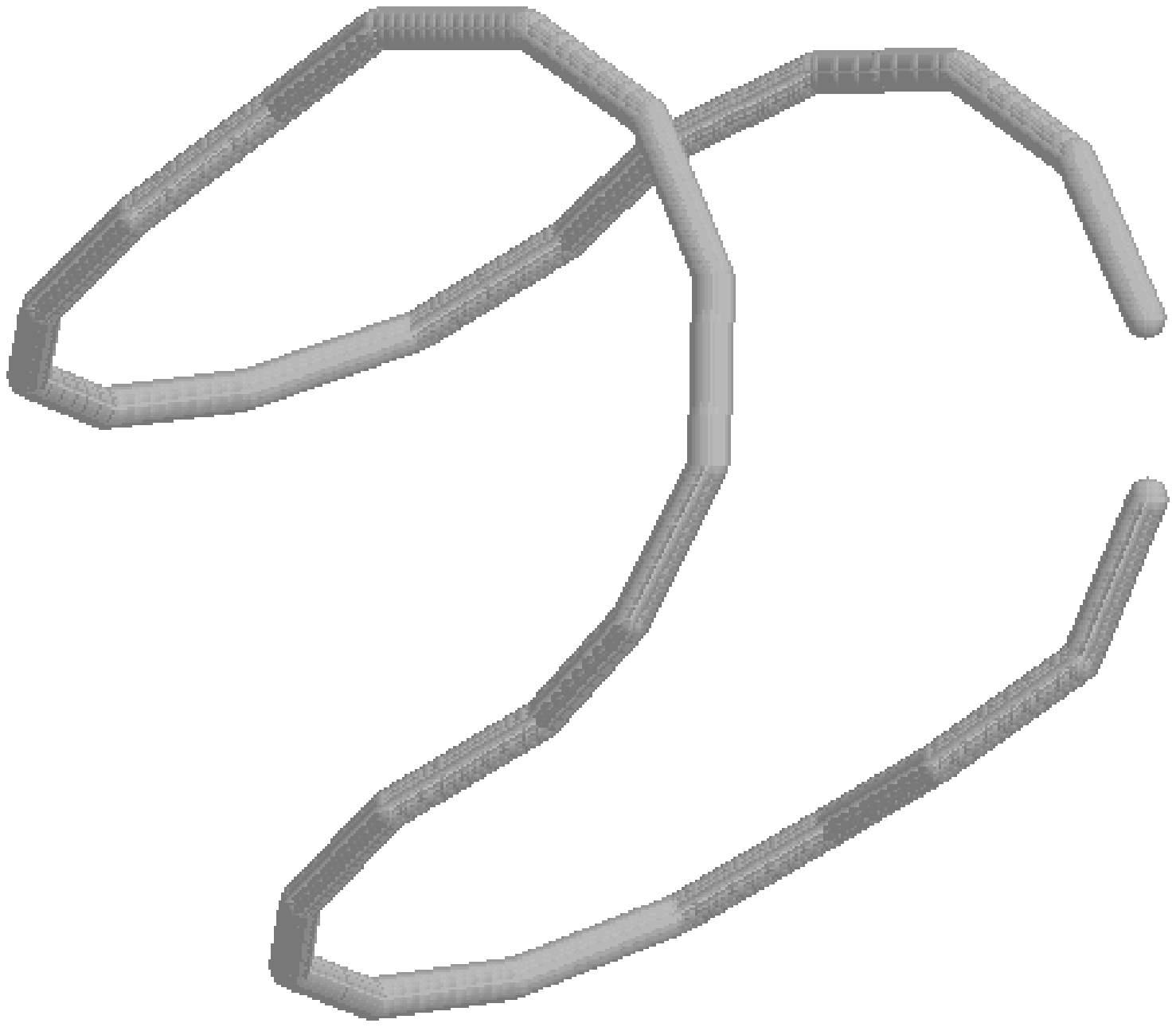}}
\\
\subfigure[]{\includegraphics[width=1.2in]{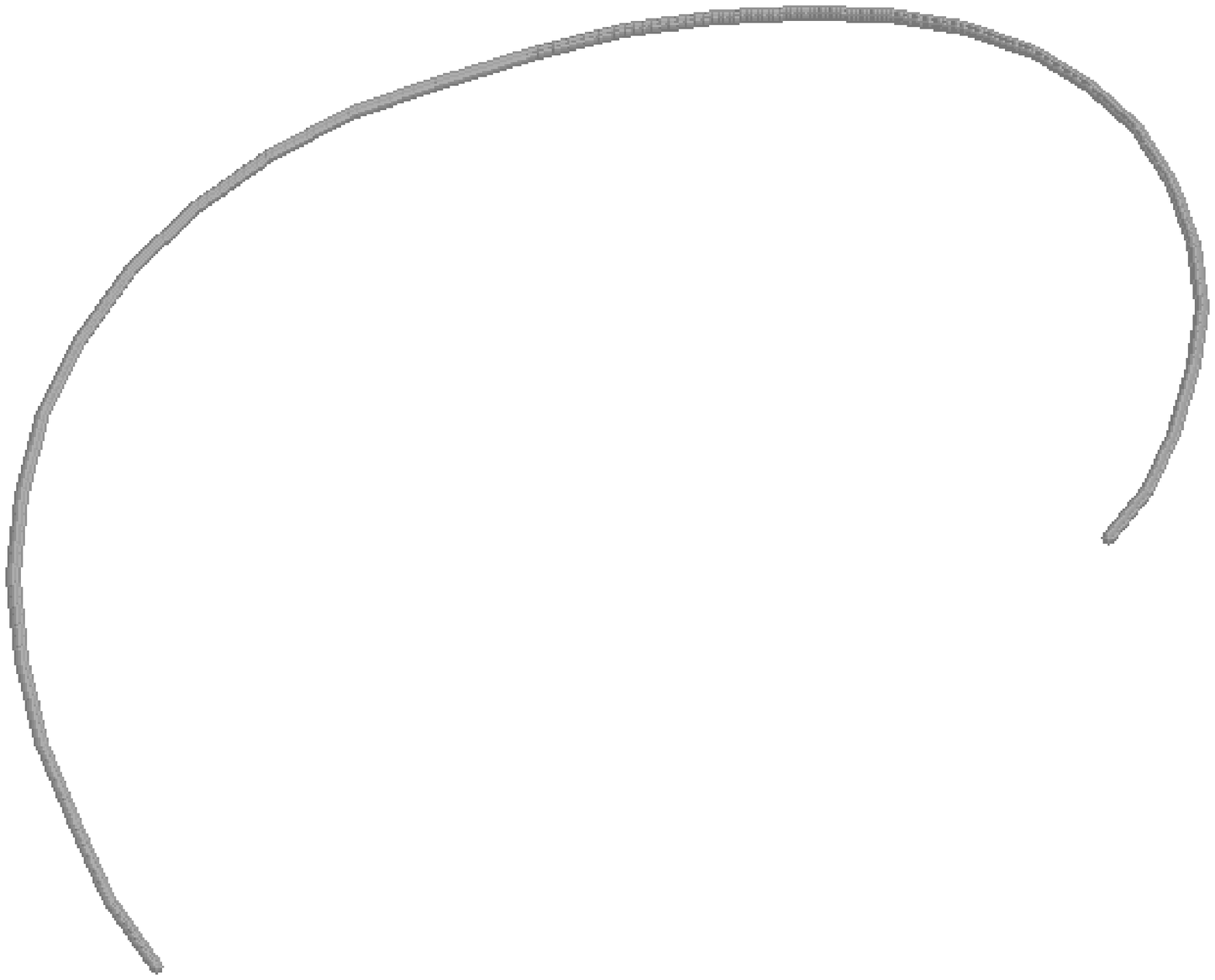}}
\hfill
\subfigure[]{\includegraphics[width=1.2in]{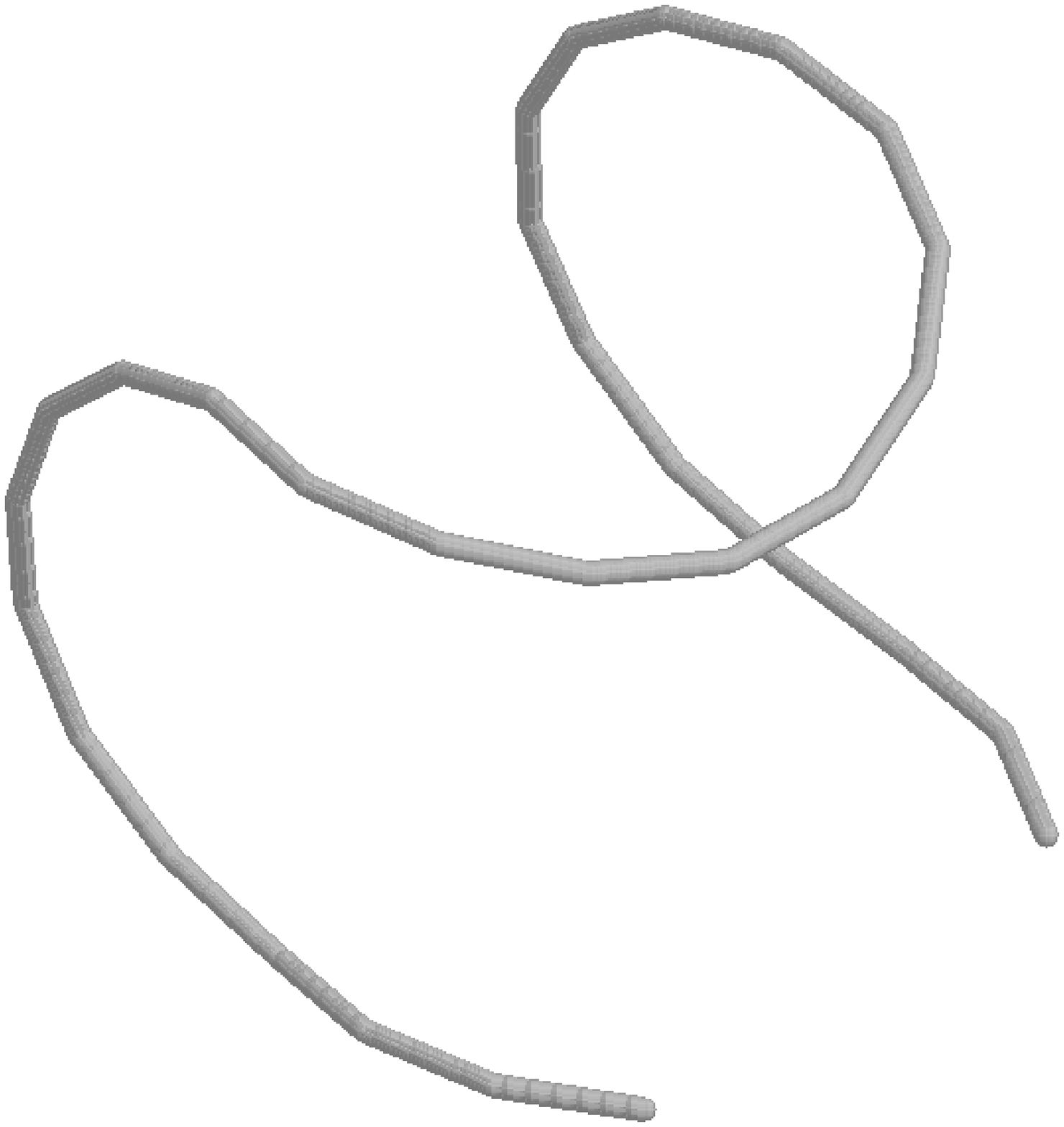}}
\hfill
\subfigure[]{\includegraphics[width=1.2in]{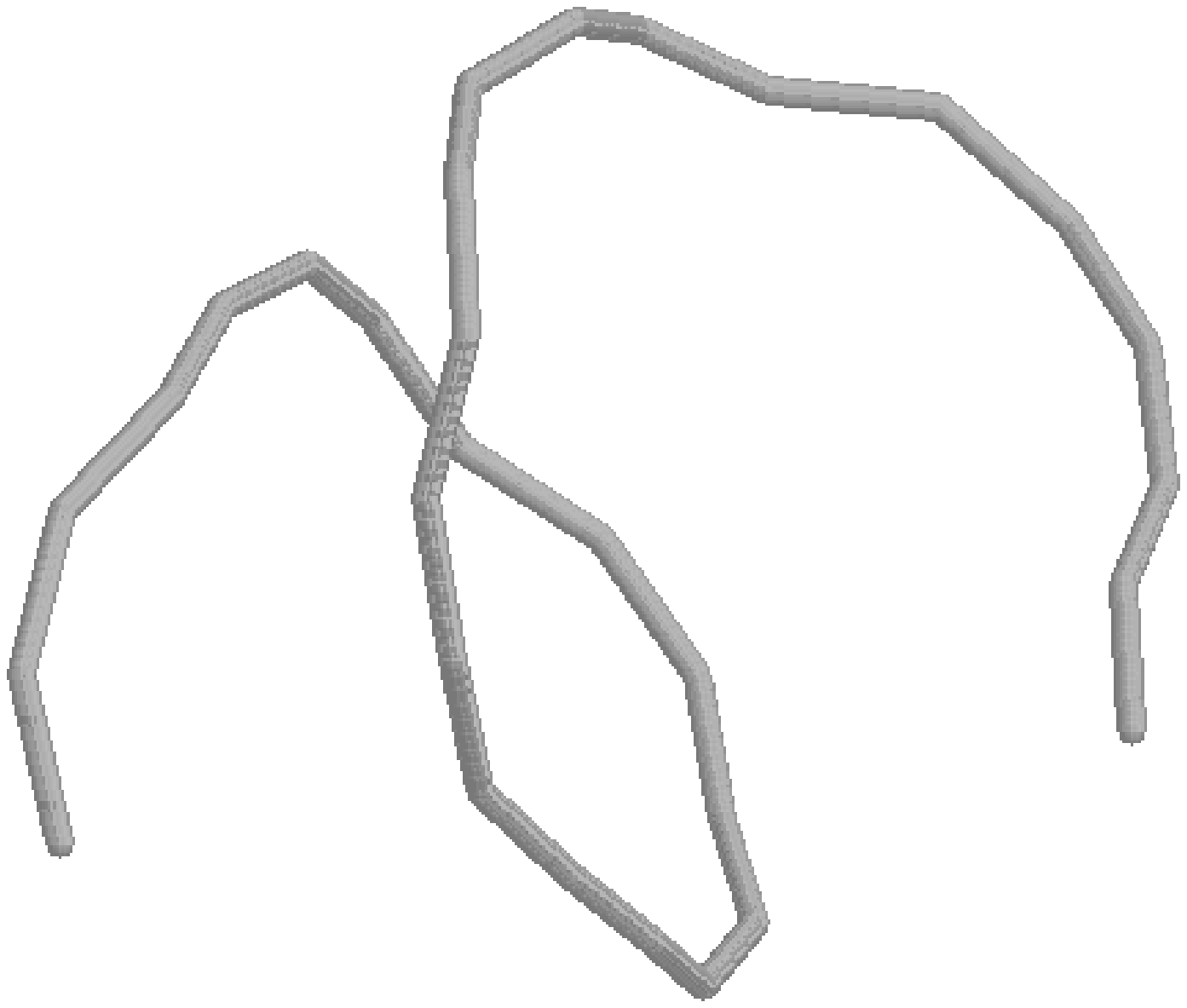}}
\caption{Examples of optimal strings. The strings in the figure were
obtained starting from a random conformation of a chain made up of $N$
equally spaced points (the spacing between neighboring points is
defined to be 1 unit) and successively distorting the chain with
pivot, crankshaft and slithering moves commonly used in stochastic
chain dynamics \cite{Sokal}. A Metropolis Monte Carlo procedure is
employed with a thermal weight, $e^{+\Delta/T}$ , where $\Delta$ is
the thickness and $T$ is a fictitious temperature set initially to a
high value such that the acceptance rate is close to 1 and then
decreased gradually to zero in several thousand steps. Self-avoidance
of the optimal string is a natural consequence of the maximization of
the thickness.  The introduction of a hard-core repulsion between
beads was found to significantly speed up convergence to the optimal
solution and avoid trapping in self-intersecting structures. We have
verified that the same values (within 1 percent) of the final
thickness of the optimal strings are obtained starting from unrelated
initial conformations. Top row: optimal shapes obtained by
constraining strings of 30 points with a radius of gyration less than
$R$. (a) $R = 6.0$, $\Delta = 6.42$ (b) $R = 4.5$, $\Delta = 3.82$ (c)
$R = 3.0$, $\Delta = 1.93$.  Bottom row: optimal shapes obtained by
confining a string of 30 points within a cube of side $L$. (d) $L =
22.0$, $\Delta = 6.11$ (e) $L = 9.5$, $\Delta = 2.3$ (f) $L = 8.1$,
$\Delta = 1.75$.}
\end{figure}

\begin{figure}
\centerline{\psfig{figure=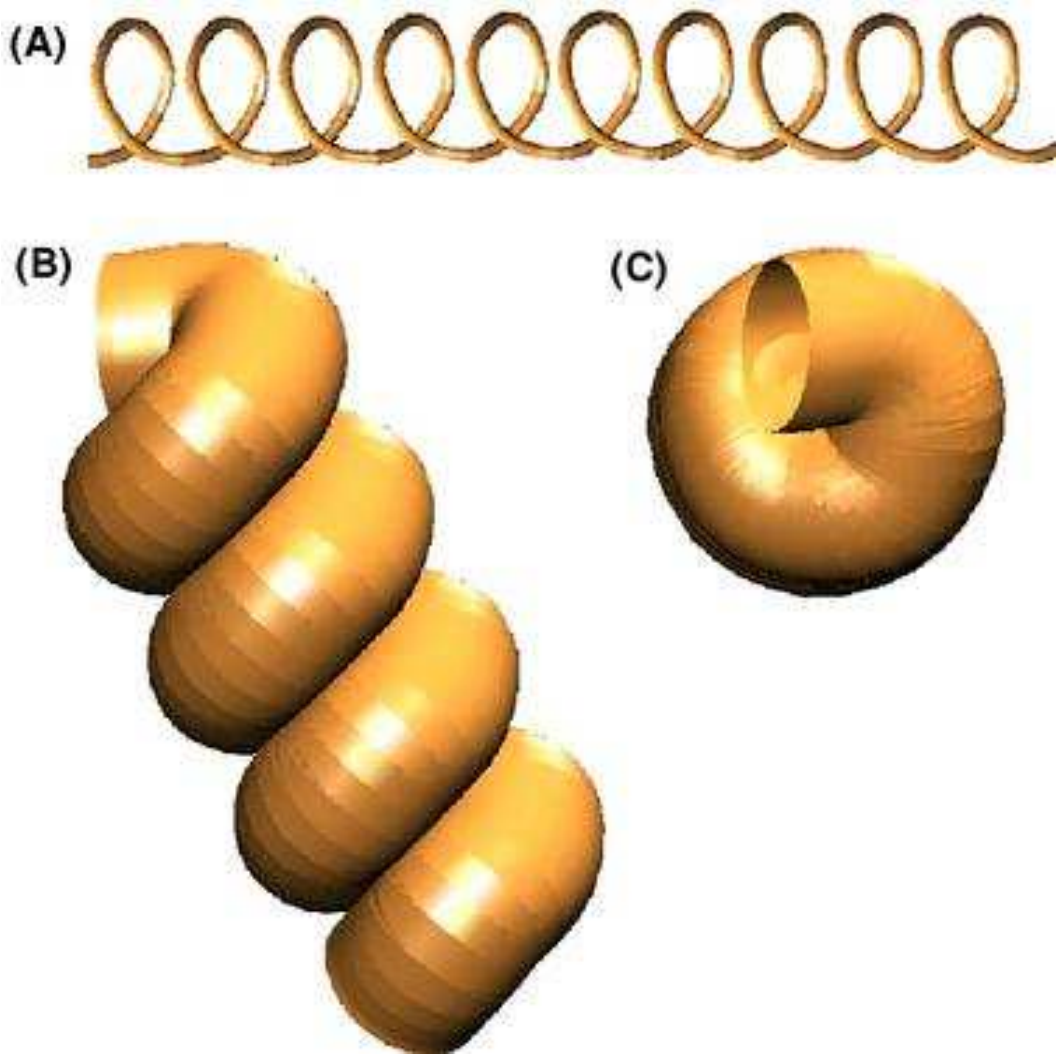,width=5.5in,height=5.5in}}
\caption{Optimal string with local constraint.  The string has 65
points with a neighbor separation of unity. The local constraint was
that each set of six consecutive beads had a radius of gyration less
than 1. The value of $f$ (see text) for this string is 0.9993. This
result is quantitatively the same for a broad class of local
constraints. Panel A shows the ``bare'' skeleton of the optimal helix
connecting the discrete beads. Panels B and C present side and top
views of the same helix inflated to its thickness.  Note that there is
no free space either between consecutive turns of the helix or in the
plane perpendicular to the helix axis.}
\end{figure}

\begin{figure}
\centerline{\psfig{figure=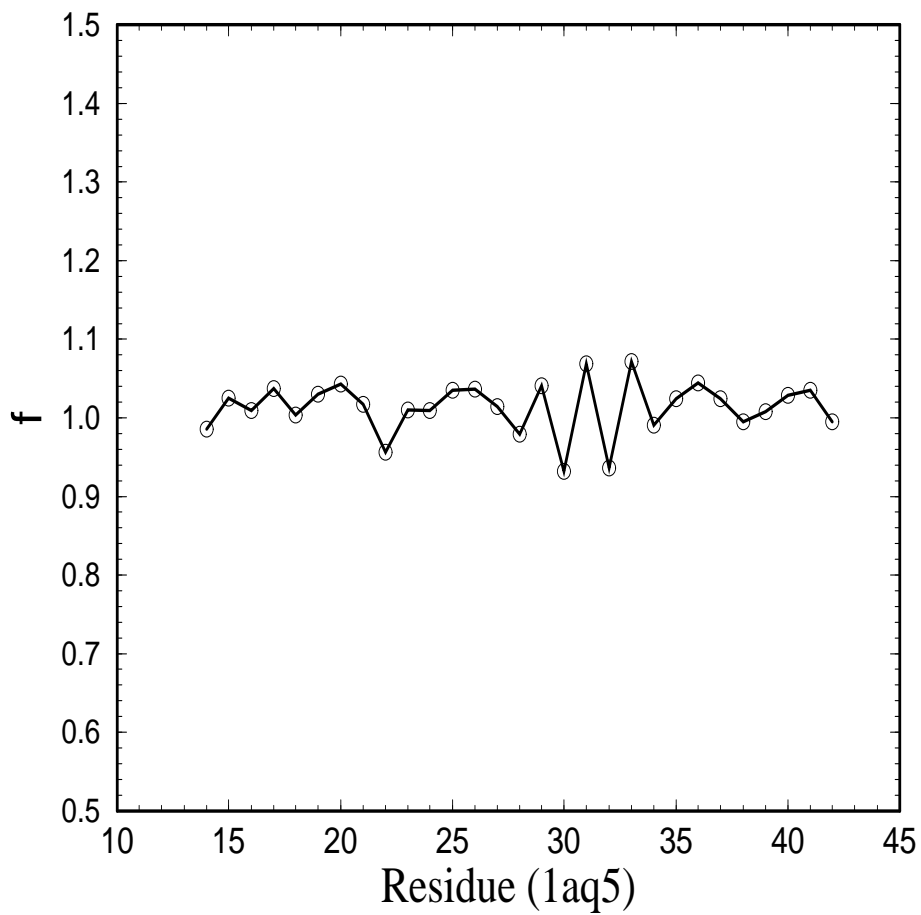,width=5.5in,height=5.5in}}
\caption{Packing of collagen helices.  Plot of $f$ values as a
function of sequence position for a single collagen helix (only
$C_\alpha$ coordinates were used to identify the protein
backbone). The same plot for each of the three collagen chains would
simply superimpose.  We considered the residues 14-42 from the
structure 1aq5 in the Protein Data Bank.}
\end{figure}

\end{document}